\documentclass[fleqn,10pt]{wlscirep}
\usepackage[utf8]{inputenc}
\usepackage[T1]{fontenc}
\usepackage{authblk}
\usepackage{lineno,hyperref}
\usepackage{graphicx}
\usepackage{color}
\usepackage{amssymb}
\usepackage{amsmath}
\usepackage{setspace}
\usepackage{babel}
\DeclareMathOperator{\sign}{sign}
\title{On the robustness of the hybrid qubit computational gates through simulated randomized benchmarking protocols}
\author[1,*]{Elena Ferraro}
\author[1,']{Marco De Michielis}
\affil{CNR-IMM Agrate Unit, Via C. Olivetti 2, 20864 Agrate Brianza (MB), Italy}
\affil[*]{elena.ferraro@mdm.imm.cnr.it}
\affil[']{marco.demichielis@mdm.imm.cnr.it}

\begin{abstract}
One of the main challenges in building a quantum processor is to characterize the environmental noise. Noise characterization can be achieved by exploiting different techniques, such as randomization where several sequences of random quantum gates are applied to the qubit under test to derive statistical characteristics about the affecting noises. A scalable and robust algorithm able to benchmark the full set of Clifford gates using randomization techniques is called randomized benchmarking. In this study, we simulated randomized benchmarking protocols in a semiconducting all-electrical three-electron double-quantum dot qubit, i.e. hybrid qubit, under different error models, that include quasi-static Gaussian and the more realistic 1/f noise model, for the input controls. The average error of specific quantum computational gates is extracted through interleaved randomized benchmarking obtained including Clifford gates between the gate of interest. It provides an estimate of the fidelity as well as theoretical bounds for the average error of the gate under test. 
\end{abstract}
\begin{document}

\flushbottom
\maketitle
\thispagestyle{empty}

\section*{Introduction}
Randomized benchmarking (RB) protocols have become an efficient tool to compare different qubit technologies and architectures \cite{Wallman-2014,Epstein-2014,McKay-2017,Carignan-2018,Wallman-2018}. Despite traditional methods of characterizing gate fidelity that involve quantum process tomography, RB possesses some advantages in terms of the robustness against the state preparation and measurement (SPAM) errors, beyond the fact that it scales efficiently with the system size, requiring fewer resources.

RB is obtained by applying sequences of feasible quantum gates of varying length, so that small errors are amplified with the sequence length leading, then, the benchmark is related to the average gate fidelity, averaged over the set of all input pure states. The quantum gates are chosen from a group of operations and Clifford gates are the most prominently considered, even if RB has also been extended to other finite groups \cite{Helsen-2019}. In practice, sequences of increasing numbers of random Clifford operations are applied to one or more qubits followed by a recovery step and a measurement. For a single qubit, the Clifford gates most commonly adopted are the $\hat{x}$, $\hat{y}$ and $\hat{z}$ rotations on the Bloch sphere. Finally, the average Clifford gate fidelity is calculated as the distance of the final state from the ideal one, as a function of the number of random Clifford operations. 

While RB results provide a significant step towards large scale benchmark of a quantum information processor, benchmarking individual gates rather than the entire set is an important instrument that could supply important information to complete the overall picture. Interleaved randomized benchmarking (IRB) is the tool to adopt in this case, in which a sequence of random Clifford gates is interleaved by the particular quantum gate (i.e. X, Z or Hadamard gate) of interest \cite{Magesan-2012,Fogarty-2015}. 
In the ideal case of perfect random gates or when the average error of all gates is depolarizing, IRB estimates the gate error perfectly. However, in reality, IRB does not give an exact characterization of the fidelity of the interleaving gate but rather provide an estimate and explicit upper and lower bounds for the gate error. These bounds give fundamental information regarding the robustness of the computational gates and the related thresholds for fault-tolerant quantum computation.

RB experiments have been performed in many different physical contexts for quantum technologies, ranging from superconducting qubits \cite{Chow-2009,McKay-2019}, trapped ions \cite{Gaebler-2012,Wright-2019} or semiconducting qubits \cite{Huang-2019,Xue-2019}, i.e. nitrogen-vacancy centers in diamond, quantum dots and donors (e.g. phosphorous) atoms in silicon. 

Realize qubit trough the confinement of electron or nuclear spins in semiconductor quantum dots is a promising approach. Much progress has been done that demonstrate the advantage of the semiconductor technology with respect to the current competing technologies in terms of fast operation times in comparison with the coherence times, prospect for scalability  and integrability with the manufacturing industry. Indeed high-fidelity control over single and multi-qubit devices based on various qubit types (i.e. single-spin qubit, double-dot singlet-triplet qubit, triple-dot exchange-only qubit, hybrid qubit) have been achieved \cite{Yoneda-2018,Zajac-2018,Nichol-2017,Barnes-2016,DiVincenzo-2000,Nakajima-2016,Shi-2012,Kim-2014,Kim-2015}.

Focusing on the hybrid qubit, the aim of the present theoretical work is to evaluate the robustness of the gate operations (X, Z and H gates) when different sources of disturbance is considered, namely Gaussian and 1/f noises. Quantitative results are reported that on one hand allow to predict the fidelity when specific error parameters are included, on the other provide a complete picture of the hybrid qubit operations in view of realization of more complex circuits, such as quantum error correction circuits.

In this paper, the analytical gate sequences, derived in Ref.\cite{Ferraro-2018}, for rotations along $\hat{x}$ and $\hat{z}$ axis for the hybrid qubit are exploited to simulate RB protocols. Gates taken from the Clifford group are used and noise is included by assuming that each operation is allowed to have some error. The quasi-static model has been employed and the control errors are modeled as random variables with Gaussian distributions featuring zero mean and standard deviation that add up to the ideal values of the corresponding control variables. Then a 1/f noise model is exploited to better take into the effects of more realistic noises including a power spectral density not constant in the frequency domain. Finally concluding remarks comparing the two noise models are given.

\section*{Results}
\subsection*{Benchmarking with quasi-static Gaussian noise}
In this Section the RB simulations for the hybrid qubit affected by a quasi-static Gaussian (QSG) noise are presented and analyzed. The hybrid qubit is an all-electrical qubit realized through confinement of three electrons in a double quantum dot \cite{Shi-2012,Ferraro-2018,Ferraro-2020}. The input controls to operate the qubit are two exchange couplings $J_1(t)$ and $J_2(t)$ between the pair of electrons belonging to different quantum dots. The exchange coupling $J$ related to the two electrons in the same dot is set by the geometry of the qubit and assumed constant. The gate sequences, derived in Ref.\cite{DeMichielis-2015, Ferraro-2018}, provide for each gate operation the sequence of exchange interaction pulses to be applied to the qubit and the corresponding duration time. 
Our sequences have control signals $J_1(t)$ and $J_2(t)$ with abrupt (ideal) switching edges. For completeness, we report in Table \ref{analytical} the analytical expressions of exchange interaction times for 2-step and 3-step sequences that realize $R_x(\theta)$ and $R_z(\theta)$, respectively, for an arbitrary angle $\theta$
\begin{table}[htbp!]
\centering
\begin{tabular}{|l|l|}
\hline
\textbf{$R_x(\theta)$}	& \textbf{$R_z(\theta)$}	\\
\hline
$t_{J_1}(\theta)=\left(\frac{n}{C}-\frac{1}{\sqrt{3}}\frac{\theta}{2\pi}\frac{1}{J_{max}}\right)h\hspace{4cm}$ & $t_{J_1}(\theta)=\frac{1}{C}\left[\frac{\theta}{\pi}A+\sign\left(\frac{2\pi}{3}-\theta\right)B\right]\frac{h}{J_{max}}\hspace{2cm}$\\
$t_{J_2}(\theta)=\left(\frac{n}{C}+\frac{1}{\sqrt{3}}\frac{\theta}{2\pi}\frac{1}{J_{max}}\right)h\hspace{4cm}$ &$t_{J_2}(\theta)=t_{J_1}(\theta)\hspace{5cm}$\\
& $t_{J}(\theta)=\left(2-\frac{\theta}{\pi}\right)\frac{h}{J_{max}}\hspace{4cm}$\\
%\hline
%$t_{tot}$=$t_{J_1}+t_{J_2}$=$\frac{2nh}{C}$ & $t_\hspace{4cm}$\\
\hline
\end{tabular}
\caption{\label{analytical}
Analytical gate sequences that realize $R_x(\theta)$ and $R_z(\theta)$ operations. In each sequence step, only one input $J_i$ is active for a $t_{J_i}$ time. Conversely, $J$ signal is kept constant during the whole sequence. In the $R_z(\theta)$ sequence, a third step of duration $t_J$ prolongs only the $J$ signal.}
\end{table}

with $A=\frac{E_z}{2}+\frac{1}{8}J_{max}$, $B=-E_z+\frac{1}{4}J_{max}$, $C=E_z+\frac{3}{4}J_{max}$, $n=\left\lceil \frac{C}{J_{max}}\frac{1}{\sqrt{3}}\frac{\theta}{2\pi} \right\rceil$ where $J_{max}=\max{(J_1)}=\max{(J_2)}$=2$J$ and $E_z$ is the Zeeman energy due to a constant global magnetic field.

The analytical sequences for the $\hat{x}$ and $\hat{z}$ single qubit gate rotations are here exploited to run the RB simulations in presence of noise. In particular the QSG model is employed and the results for different values of the amplitude and time errors of the input controls are shown, then the IRB for X, Z and H gates are simulated and compared to RB. Gaussian noise model represents a reasonable approximation for sufficiently short qubit operations, it can be classified as a Markovian noise that is memoryless and then history independent. The hypothesis is indeed that noise fluctuations do not depend on the length of the sequence.

The RB sequence is built starting from random gate operations chosen uniformly from the Clifford group on n-qubits, then a computed reversal element is included, that in the ideal case should return the qubits to the initial state. 

We focus on single-qubit Clifford operations generated through the native gate set: $\{I, X(\pi), Y(\pm\pi), X(\pm\pi/2), Y(\pm\pi/2)\}$, where an arbitrary Y rotation around an angle $\theta$ is obtained composing X and Z gate for which we have the analytical expressions, i.e. $Y(\theta)=Z(-\pi/2)X(\theta)Z(\pi/2)$. 
The signal pulses to implement the gates have a maximum exchange interaction amplitude $J_{max}$=1 $\mu$eV and a minimum duration $t_{min}$= 100 ps, resulting in total gate times given in Table \ref{gatetime}: 

\begin{table}[htbp!]
\centering
\begin{tabular}{|l|l|}
\hline
Gate & Time\\
\hline
$I$ & 5.29 ns\\ 
$X(\pm\pi)$ &  2.80 ns\\ 
$X(\pm\pi/2)$ & 1.55 ns\\  
$Z(\pi/2)$ & 16.04 ns\\
$Z(-\pi/2)$ & 16.12 ns\\ 
$Y(\pm\pi)$ & 34.96 ns\\
$Y(\pm\pi/2)$ & 33.71 ns\\
\hline
\end{tabular}
\caption{\label{gatetime}
Gate times for I, X, Y an Z operations.}
\end{table}

In practice, for each values of $N$ that represent the number of the gate operations composing the sequence, we choose also the number of sequences to simulate that we denote by $N_{seq}$. Each such sequence contains $N$ random elements chosen uniformly from the Clifford group and the $N$+1 element that is defined as the reversed. 

Moreover, we assume that each operation is allowed to have some error that is included as an additional disturbed sequence in which the control errors are modeled as random variables with Gaussian distributions featuring zero mean and standard deviation $\sigma$. For the time interval, the error with standard deviation $\sigma_t$, is added up to the ideal sequence time step whereas the additional error on the exchange interaction $J$ is obtained by multiplying a normalized error with standard deviation $\sigma_J/J_{max}$ to the different control inputs $J$. We repeat this procedure $N_{rep}$ times and then we make the average. 

Finally the results obtained for the averaged sequence fidelity are fitted to the model function
\begin{equation}
f=A+Bp^N 
\end{equation}
where A, B and p are the fit parameters and $N$ is the number of operations composing the sequence.  

To choose the initial condition there are alternative routes. The simpler one is to choose as initial condition a simple pure initial state, for example $|\psi(0)\rangle=|1\rangle$. However more information can be extracted considering an average over the set of all input pure states on the Bloch sphere. A good compromise is to consider an average of states on the Bloch sphere varying appropriately the angles $\theta$ and $\phi$. In the following we study the initial condition obtained as an average over the six states: $\{|0\rangle, \frac{1}{\sqrt{2}}(|0\rangle\pm|1\rangle),\frac{1}{\sqrt{2}}(|0\rangle\pm i|1\rangle),|1\rangle\}$. We have verified that there are no significant deviations considering an average state obtained with larger number of states on the Bloch sphere.

\begin{figure}[htbp!]
\centering
a)\includegraphics[width=0.45\textwidth]{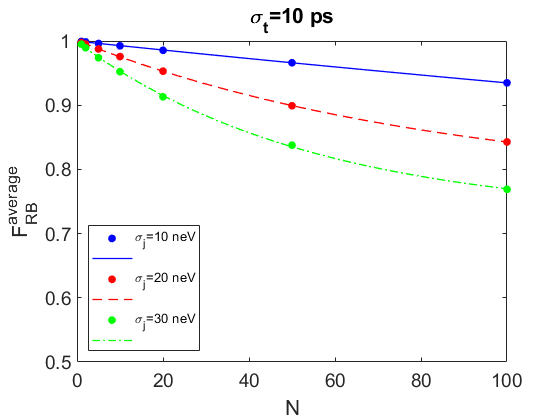}\quad b)\includegraphics[width=0.45\textwidth]{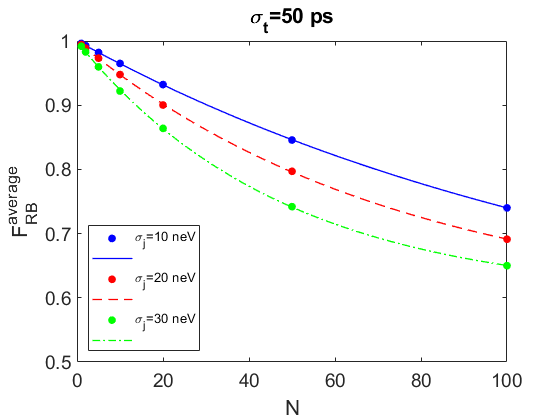}\\
c)\includegraphics[width=0.45\textwidth]{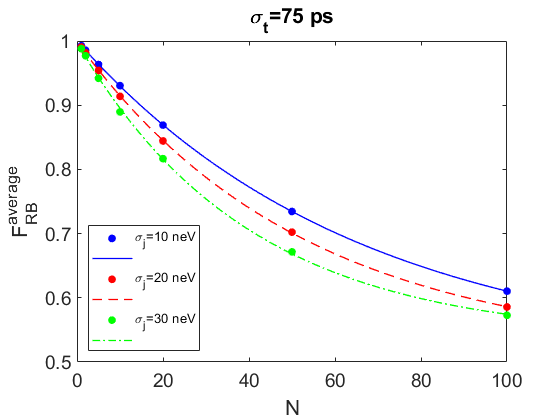}\quad d)\includegraphics[width=0.45\textwidth]{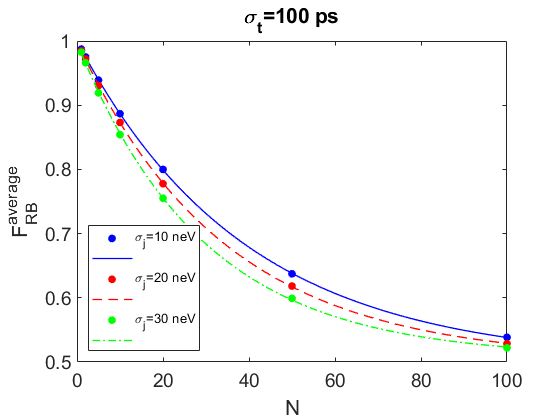}\\
\caption{Average fidelity RB with $N_{seq}$=800 and $N_{rep}$=10 in presence of QSG noise model for different values of $\sigma_j$= 10, 20, 30 neV and $\sigma_t$= 10 ps (a), 50 ps (b), 75 ps (c), 100 ps (d). The dots represent the result of the numerical simulation, the lines represent the fit result.}\label{RB_ave_800}
\end{figure}

Figure \ref{RB_ave_800} shows the average fidelity RB with $N_{seq}$=800 and $N_{rep}$=10 for different values of the standard deviation $\sigma_j$= 10, 20, 30 neV for the amplitude of the input control $J$ and $\sigma_t$= 10 ps (a), 50 ps (b), 75 ps (c), 100 ps (d) for the the standard deviations related to the time interval error of the applied pulse. The choice of the parameters used in the following simulations is guided by the realistic values for $J_{max}$ and the gate times in the hybrid qubit experiments. If the maximum exchange coupling is fixed to 1 $\mu$eV, considering standard deviations $\sigma_j$= 10, 20, 30 neV correspond to include an error of 1\%, 2\% and 3\% respectively. Similarly for the gate times, that are in the range of hundreds of ns, we consider error standard deviations compatible with state-of-the-art experimental apparatus \cite{DatasheetMWgenerator}. The dots represent the result of the numerical simulation, the lines represent the fit result. The behaviour of $F_{RB}^{average}$ as a function of the number of the gate operations $N$ shows the typical exponential decay, with a reduction of the fidelity when $\sigma_t$ and $\sigma_j$ are increased. $\sigma_j$ affects strongly the fidelity for small $\sigma_t$ values (see panel a). When $\sigma_t$ is increased to 100 ps (panel d) the different fidelity curves are very close, with $F_{RB}^{average}$ almost reaching 0.5 at $N=100$ for $\sigma_j$=30 neV.     

From the fit it is possible to extract interesting information calculating an average error rate, also called Error per Clifford (EPC) that in the case of one qubit is defined as EPC=$\frac{1}{2}(1-p)$. The fit parameters A and B instead absorb SPAM errors as well as an edge effect from the error on the final gate. 

\begin{figure}[htbp!]
\centering
\includegraphics[width=0.5\textwidth]{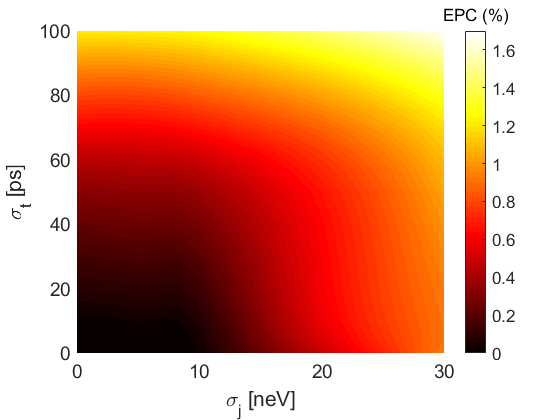}
\caption{EPC with $N_{seq}$=800 and $N_{rep}$=10 in presence of a QSG noise model.}\label{EPC_ave}
\end{figure}

Figure \ref{EPC_ave} shows a two-dimensional colored map of EPC expressed in \% as a function of the standard deviations $\sigma_j$ and $\sigma_t$. The greater is the error on the input controls in amplitude and in time, the greater is the EPC, that for the range of parameters considered remains below 2\%.

\subsubsection*{Interleaved Randomized Benchmarking}
Any arbitrary SU(2) gate U can be constructed combining $\hat{x}$ and $\hat{z}$ rotations on the Bloch sphere. It is possible to demonstrate that in general, up to a global phase, \cite{McKay-2017}
\begin{equation}\label{U}
U(\phi,\theta,\lambda)=Z_{\phi}X_{\theta}Z_{\lambda}.
\end{equation}

\begin{figure}[htbp!]
\centering
a)\includegraphics[width=0.45\textwidth]{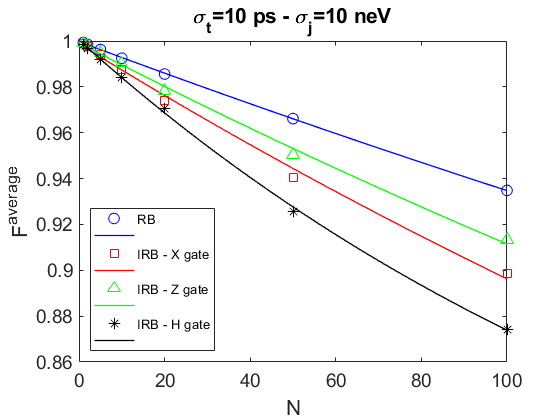}\quad b)\includegraphics[width=0.45\textwidth]{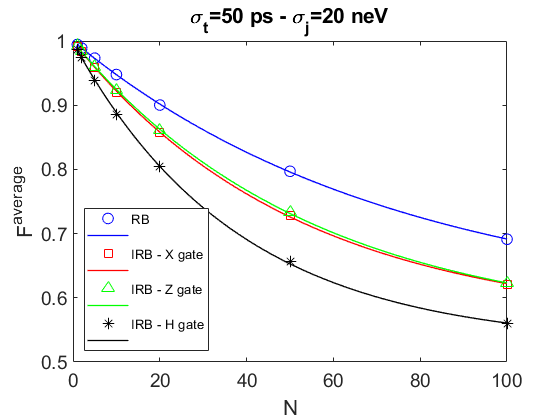}\\
\caption{Comparison RB - IRB for X, Z and H gates with $N_{seq}$=800 and $N_{rep}$=10 in presence of QSG noise model for different values of $(\sigma_t,\sigma_j)$ = 10 ps, 10 neV (a); 50 ps, 20 neV (b). The markers represent the result of the numerical simulation, the lines represent the fit result.}\label{Comparison_IRB}
\end{figure}

Figure \ref{Comparison_IRB} shows a comparison between RB and IRB for different gate operations belonging to the Clifford group, that are X, Z and H gates, the latter is obtained inserting in Eq. (\ref{U}) the parameters $\phi=\theta=\lambda=\pi/2$. Total operation times for X, Z and H gates are $t_X$=2.80 ns, $t_Z$= 16.00 ns and $t_H$=33.63 ns, respectively. Each panel shows a different choice for the standard deviations: $(\sigma_t,\sigma_j)$ = 10 ps, 10 neV (a); 50 ps, 20 neV (b).

$F^{average}$ shows a decay when $N$ is increased and is reduced as $\sigma_j$ and $\sigma_t$ are increased. All the gates have an $F^{average}$ lower than the RB fidelity. In addition the X gate fidelity depends also on the entity of the error considered. We observe indeed that while in the low noise case (Fig. \ref{Comparison_IRB}a) it is lower than the $F^{average}$ for the Z gate, in the high noise case where also $\sigma_j$ grows (Fig. \ref{Comparison_IRB}b) the two fidelities become comparable. The X gate is the faster of the three gates here studied, therefore we may argue that the error on the time step quantified by $\sigma_t$ has a greater weight in the IRB-X gate evaluation with respect to the other gates studied. The H gate, that is obtained with the longer sequence, shows in both cases the lower fidelity. 

In Tables \ref{parameters1} and \ref{parameters2} we report the calculated gate errors of the associated noise operator and the relative bounds in the two cases considered in Figs. \ref{Comparison_IRB}a and  \ref{Comparison_IRB}b respectively. The gate error is defined as $\varepsilon=\frac{1}{2}\left(1-\frac{p^i}{p}\right)$, where $p^i$ is the fit parameter corresponding to the gate under study ($i\equiv$ X, Z and H) and $p$ is the RB fit parameter and it lies in the interval [$\varepsilon$-E, $\varepsilon$+E] where \cite{Magesan-2012}:

\begin{equation}
E=\min\left\{\frac{1}{2}\left[\left|p-\frac{p^i}{p}\right|+(1-p)\right],\left[\frac{3}{2}\frac{1-p}{p}+\frac{4\sqrt{3}\sqrt{1-p}}{p}\right]\right\}.
\end{equation}

\begin{table}[htbp!]
\centering
\begin{tabular}{|l|l|l|l|}
\hline
Gate & $p^i$ & $\varepsilon$ & Bounds\\
\hline
X & $0.9976\pm 0.0020$ & $0.0005\pm 0.0012$ & [0, 0.0014]\\ 
Z &  $0.9980\pm 0.0020$ & $0.0003\pm 0.0010$ & [0, 0.0014]\\ 
H &  $0.9940\pm 0.0009$ & $0.0023\pm 0.0005$ &[0, 0.0046]\\
\hline
\end{tabular}
\caption{\label{parameters1}
Gate error $\varepsilon$ for each noise operator and the relative bounds in presence of QSG noise model. RB parameters: $p=0.9986\pm 0.0004$, EPC=$0.0007\pm 0.0002$. The standard deviations are $\sigma_t$=10 ps, $\sigma_j$=10 neV.}
\end{table}

\begin{table}[htbp!]
\centering
\begin{tabular}{|l|l|l|l|}
\hline
Gate & $p^i$ & $\varepsilon$ & Bounds\\
\hline
X & $0.9808\pm 0.0004$ & $0.0030\pm 0.0003$ & [0, 0.0133]\\ 
Z &  $0.9821\pm 0.0004$ & $0.0024\pm 0.0002$ & [0, 0.0133]\\ 
H &  $0.9739\pm 0.0007$ & $0.0065\pm 0.0004$ &[0, 0.0133]\\ 
\hline
\end{tabular}
\caption{\label{parameters2}
Gate error $\varepsilon$ for each noise operator and the relative bounds in presence of QSG noise model. RB parameters: $p=0.9867\pm 0.0003$, EPC=$0.0066\pm 0.0001$. The standard deviations are $\sigma_t$=50 ps, $\sigma_j$=20 neV.}
\end{table}

The H gate presents higher gate error values with respect to the other considered gates. This can be understood by looking at its longer sequence, as shown in Eq. \ref{U}, than the other gate sequences. As a result, the IRB sequences for the H gate in Figure \ref{Comparison_IRB} have a much longer mean duration than the RB sequences interleaved with X and Z ones, leading to higher gate errors. 

\subsection*{Benchmarking with 1/f noise}
The RB analysis on the hybrid qubit is enriched with the inclusion of the frequency dependence of the noise that is here modeled to be of the 1/f type \cite{Epstein-2014,Zhang-2017,Yang-2019PRA}. On the contrary of the QSG noise model, 1/f model can be classified as a non-Markovian noise that is history dependent, in such a way that the noise at one moment depends on the duration of the previous gates in the sequence. 1/f noise is ubiquitous in nature and is present in several physical implementations of qubits \cite{Paladino-2014}.

The definition of the 1/f model is based on the shape of the Power Spectral Density (PSD) that is given by $S_j(\omega)$ = $A_j/(\omega t_0)$, where $A_j$ is the amplitude, that does not depend on $\omega$ and $t_0$ is the time unit. In order to fairly compare the fidelity obtained with the 1/f noise with respect to the QSG one, we impose the equivalence between the power of the noises. In correspondence to $\sigma_j$= 10, 20, 30 neV of the QSG noise, the 1/f noise amplitudes are given by $A_j=\pi t_0 h\left(\frac{\sigma_j}{J_{max}t_0}\right)^2/\ln\left(\frac{f_{max}}{f_{min}}\right)$= 1.0644, 4.2577, 9.5799 neV, respectively, where $f_{min}$= 50 kHz and $f_{max}$= 10 GHz are the low and high frequency cutoffs and are taken from Ref. \cite{Zhang-2017}, and $t_0=1/f_{max}$.

\begin{figure}[htbp!]
\centering
a)\includegraphics[width=0.45\textwidth]{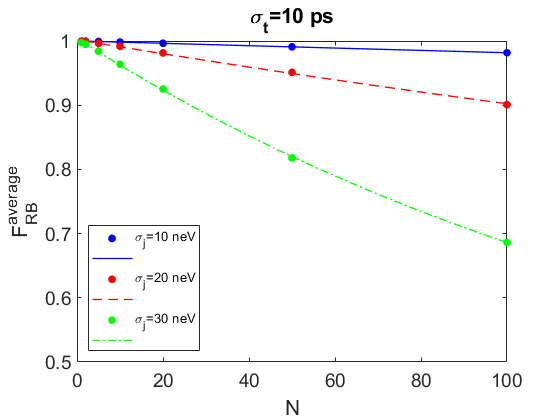}\quad b)\includegraphics[width=0.45\textwidth]{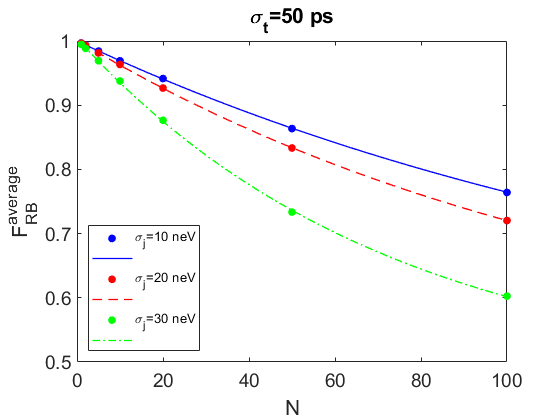}\\
c)\includegraphics[width=0.45\textwidth]{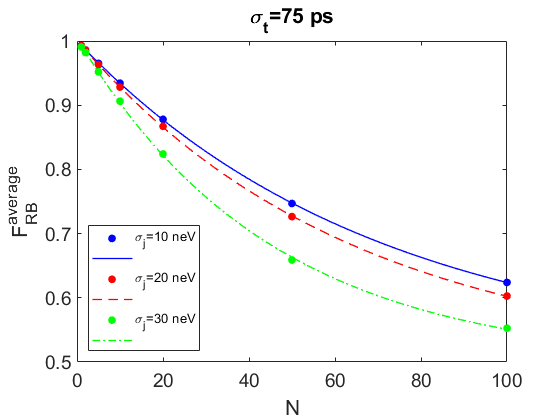}\quad d)\includegraphics[width=0.45\textwidth]{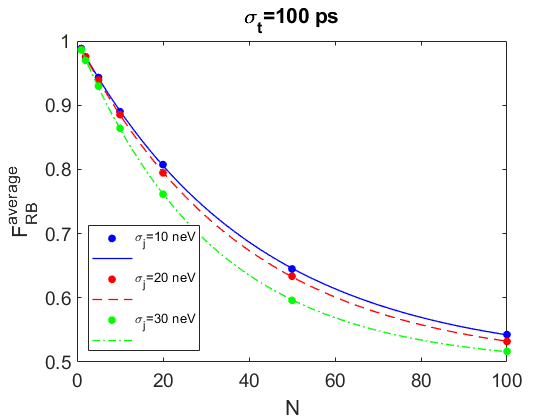}\\
\caption{Average fidelity RB with $N_{seq}$=800 and $N_{rep}$=10 in presence of 1/f noise model for different values of the amplitude $A_j$ labeled with the corresponding values of the equivalent QSG noise with standard deviation $\sigma_j$=10, 20, 30 neV. The values for $\sigma_t$ are 10 ps (a), 50 ps (b), 75 ps (c), 100 ps (d).}\label{1overf}
\end{figure}

Figure \ref{1overf} shows the average fidelity RB with $N_{seq}$=800 and $N_{rep}$=10 in presence of 1/f noise model for different values of the amplitude $A_j$ that here is labeled with the corresponding values of an equivalent QSG noise with standard deviation $\sigma_j$= 10, 20, 30 neV. The behaviour of $F_{RB}^{average}$ as a function of the number of the gate operations $N$ shows the typical exponential decay. When $\sigma_t$ and $\sigma_j$ are increased the RB fidelity decreases.

\begin{figure}[htbp!]
\centering
\includegraphics[width=0.5\textwidth]{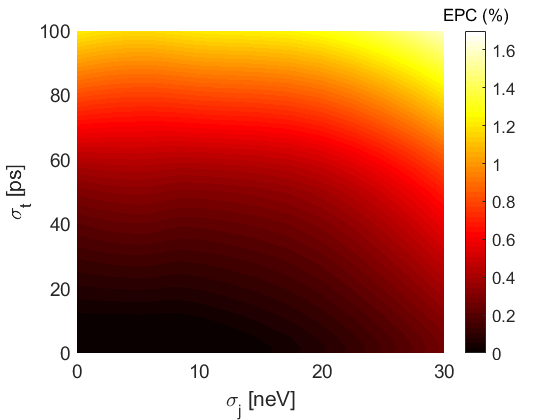}
\caption{EPC with $N_{seq}$=800 and $N_{rep}$=10 in presence of 1/f noise model.}\label{EPC_1overf}
\end{figure}

Figure \ref{EPC_1overf} shows how the EPC changes as $\sigma_j$ and $\sigma_t$ are modified. The EPC dependence on $\sigma_j$ is less pronounced with respect to the QSG noise with the same power, reported in Figure \ref{EPC_ave}.

\subsubsection*{Interleaved Randomized Benchmarking}
\begin{figure}[htbp!]
\centering
a)\includegraphics[width=0.45\textwidth]{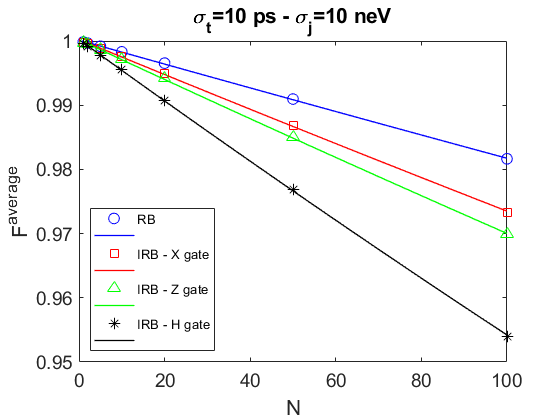}\quad b)\includegraphics[width=0.45\textwidth]{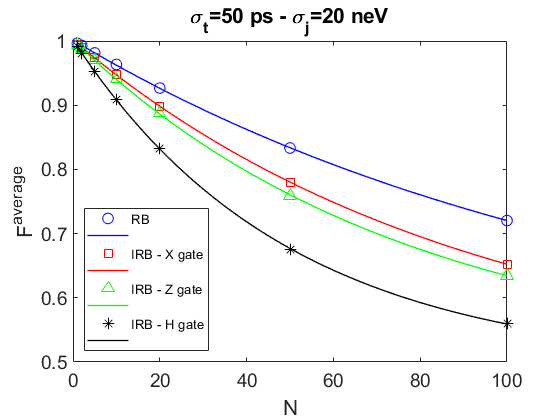}\\
\caption{Comparison RB - IRB for X, Z and H gates with $N_{seq}$=800 and $N_{rep}$=10 in presence of 1/f noise model for different values of $(\sigma_t, \sigma_j)$ = 10 ps, 10 neV (a); 50 ps, 20 neV (b).}\label{Comparison_IRB_1overf}
\end{figure}

IRB analysis is here implemented including the 1/f noise model. Figure \ref{Comparison_IRB_1overf} shows a comparison between RB and IRB for X, Z and H gates. As for the QSG case, each panel shows a different choice for the error parameters $(\sigma_t,\sigma_j)$ = 10 ps, 10 neV (a) and 50 ps, 20 neV (b). $F^{average}$ shows a similar qualitatively behaviour of the one derived with the QSG model (see Figure \ref{Comparison_IRB}) where an exponential decay when $N$ is increased is observed. Analogously, for each gate the corresponding $F^{average}$ is lower than the RB fidelity deteriorating as $\sigma_j$ and $\sigma_t$ are increased. We observe that $F^{average}$ for the X gate results higher than the one of the Z gate that presents an higher fidelity with respect to the H gate in both cases considered. 

The gate error of the associated noise operator and the relative bounds in both cases studied are reported in Tables \ref{parameters1_1overf} and \ref{parameters2_1overf}. 

\begin{table}[htbp!]
\centering
\begin{tabular}{|l|l|l|l|}
\hline
Gate & $p^i$ & $\varepsilon$ & Bounds\\
\hline
X & $0.9994\pm 0.0007$ & $0.0001\pm 0.0004$ & [0, 0.0004]\\ 
Z &  $0.9994\pm 0.0005$ & $0.0001\pm 0.0004$ & [0, 0.0004]\\ 
H &  $0.9990\pm 0.0004$ & $0.0003\pm 0.0003$ &[0, 0.0006]\\ 
\hline
\end{tabular}
\caption{\label{parameters1_1overf}
Gate error $\varepsilon$ for each noise operator and the relative bounds in presence of 1/f noise model. RB parameters: $p=0.9996\pm 0.0006$, EPC=$0.0002\pm 0.0003$. The error parameters are $\sigma_t$=10 ps, $\sigma_j$=10 neV.}
\end{table}

\begin{table}[htbp!]
\centering
\begin{tabular}{|l|l|l|l|}
\hline
Gate & $p^i$ & $\varepsilon$ & Bounds\\
\hline
X & $0.9891\pm 0.0002$ & $0.0016\pm 0.0001$ & [0, 0.0076]\\ 
Z &  $0.9871\pm 0.0002$ & $0.0026\pm 0.0002$ & [0, 0.0076]\\ 
H &  $0.9799\pm 0.0002$ & $0.0062\pm 0.0002$ &[0, 0.0124]\\ 
\hline
\end{tabular}
\caption{\label{parameters2_1overf}
Gate error $\varepsilon$ for each noise operator and the relative bounds in presence of 1/f noise model. RB parameters: $p=0.9922\pm 0.0002$, EPC=$0.0038\pm 0.0001$. The error parameters are $\sigma_t$=50 ps, $\sigma_j$=20 neV.}
\end{table}

The H gate presents larger values for the gate error with respect the other gates considered. This is due to its longer sequence duration than the other gate sequence lengths.

\subsection*{RB comparison between QSG and 1/f noise models}
\begin{figure}[htbp!]
\centering
\includegraphics[width=0.5\textwidth]{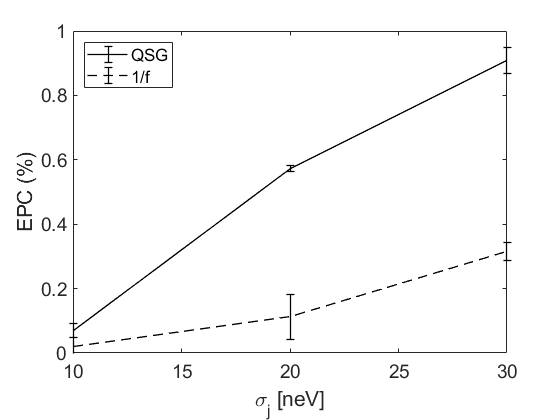}
\caption{EPC comparison for QSG and 1/f noise models in correspondence to $\sigma_t$ = 10 ps.}\label{Comparison_QSGN_1overf}
\end{figure}

The two noise models are here quantitatively compared. Figure \ref{Comparison_QSGN_1overf} shows a comparison between the EPC obtained by RB with the QSG noise model and the 1/f noise one as a function of $\sigma_j$. The standard deviation $\sigma_t$ = 10 ps has been chosen in order to maximize the EPC modulation due to $\sigma_j$. Using the same noise power, the EPC for the 1/f model is significantly lower (up to 0.6\%) than that obtained with the QSG noise model for every $\sigma_j$ considered in this study.

\begin{figure}[htbp!]
\centering
\includegraphics[width=0.5\textwidth]{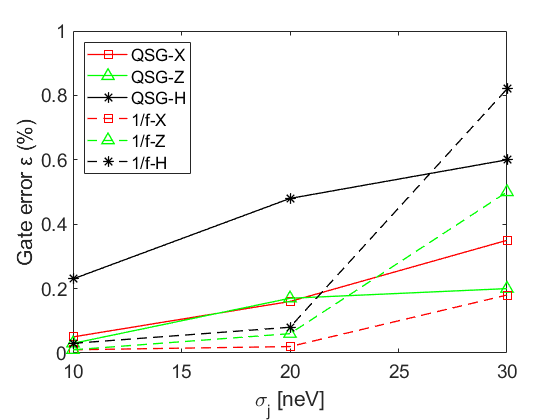}
\caption{Gate error $\varepsilon$ comparison for each noise operator for QSG and 1/f noise models in correspondence to $\sigma_t$ = 10 ps.}\label{Comparison_QSGN_1overf_eps}
\end{figure}

Figure \ref{Comparison_QSGN_1overf_eps} shows a comparison between the gate error $\epsilon$ obtained by IRB with the QSG noise model and the 1/f noise one as a function of $\sigma_j$ with $\sigma_t$ = 10 ps for X, Z and H gates. Imposing the same noise power, we observe different behaviour depending on the values of $\sigma_j$ considered. For $\sigma_j<$ 20 neV the gate errors $\epsilon$ for X, Z and H gates calculated in correspondence to the 1/f model are lower than the corresponding cases for the QSG noise model. Conversely, when $\sigma_j>$ 20 neV the behaviour is the opposite with values for the gate error higher in correspondence to the 1/f model. An exception is represented by the X gate, the faster gate with respect to the other studied, whose gate error calculated with the 1/f model stay below the one calculated with the QSG model also for large values of $\sigma_j$. Moreover, we observe that for the X gate, $\epsilon$ values are less sensitive to $\sigma_j$ modulation when compared to the other gate results.

\section*{Discussion}
Randomized Benchmarking is a powerful tool to characterize the noise and compare different qubit technologies. We perform the first comprehensive  study of the single qubit gate fidelity in the hybrid qubit starting from analytical gate sequences for $\hat{x}$ and $\hat{z}$ rotations on the Bloch sphere by means of simulations of RB protocols. The quasi-static Gaussian and 1/f noise models for the input controls are considered. 
First, by exploiting the QSG noise model, we obtained a maximum value for EPC equal to 1.67\% for the entire range of error parameters considered. Then individual gates, that are X, Z and Hadamard gates, are benchmarked through simulation of Interleaved Randomized Benchmarking protocols from which an estimation of the gate error is extracted giving explicit upper and lower bounds. We have found for the QSG model in correspondence to two different noise configurations ($\sigma_t$=10 ps, $\sigma_j$=10 neV) and ($\sigma_t$=50 ps, $\sigma_j$=20 neV), that X and Z gate error upper bounds are almost identical, ranging from 0.14\% to 1.33\%, and Hadamard gate ones ranging from 0.46\% to 1.33\%. The latter gate, obtained composing rotations along $\hat{x}$ and $\hat{z}$ axis, is the most sensitive to noise. 
Then, for the 1/f noise model, the RB simulations gave a maximum EPC value equal to 1.59\% whereas IRB simulated protocols, for the same two noise configurations considered in the QSG model, resulted in X and Z gate error upper bounds ranging from 0.04\% to 0.76\% and Hadamard ones from 0.06\% to 1.24\%. 
By comparing the effects of the two noise models with the same noise power, we found that QSG noise overestimates the fidelity reduction with respect to that obtained with the 1/f noise, affecting heavily the gates with longer sequences than those with shorter ones. We observe also that the obtained results are very sensitive to the value of the noise parameter $\sigma_j$ that, when large, affects in a stronger way the IRB fidelity calculated through the 1/f noise model. The results of the 1/f model in common experimental conditions of $\sigma_t$=10 ps and $\sigma_j$=20 neV (i.e. 2\% of $J_{max}$) show that gate errors in the 0.1\% range can be achieved, enabling the implementation of the most advanced quantum error correction circuits.   

\section*{Methods}
Following Ref. \cite{Yang-2016} we generated the 1/f noise in the frequency domain as:
\begin{equation}\label{eq:S1overf}
n(\omega)=m(\omega)^{-1/2} e^{i\phi(\omega)}
\end{equation}
where $m(\omega)$ is generated from a standard Gaussian white process and the phase factor $\phi(\omega)$ is taken from a uniform distribution between 0 and $2\pi$. The final noise in the time domain is obtained by applying an inverse Fourier transform to the above equation discretized in the frequency domain and then by multiplying the result by the noise amplitude $A_j$.

\bibliography{Ref}
\section*{Author contributions statement}
E.F. and M.D.M. contributed equally to conceive and design the theoretical study, conduct the numerical calculation and analyze the results, write and review the manuscript.
\end{document}